# New developments in the geometry and topology of 3D head-direction coding

*Jean Laurens*

*Ernst Strüngmann Institute of the Max Planck Society, Frankfurt, Germany*

## Abstract

The head-direction (HD) system is a neuronal compass that represents head orientation in 3D. Its encoding of 3D head orientation can be decomposed into a multiplexed representation of 2D head tilt relative to gravity and 1D head orientation in the horizontal plane (azimuth). Defining azimuth during 3D motion is not trivial, and recent studies have indicated that the HD system implements a so-called dual-axis/tilted-azimuth (DA/TA) rule, a representation that is well supported experimentally but remains mathematically and mechanistically difficult to interpret. Here, I introduce a new formulation of DA/TA based directly on the elements of a rotation matrix describing 3D head orientation. In this formulation, TA is represented by a 2D vector, whose direction encode TA and whose norm depends on tilt, consistent with a tilt-dependent gain of HD tuning reported by experimental studies. I provide a geometrical interpretation of this formulation and use it to characterize the topology of the resulting representation. More generally, I introduce a framework in which head orientation is represented by a rotation matrix and neuronal responses arise from linear combinations of its elements followed by nonlinear transformations. Within this framework, the linear stage specifies which aspects of 3D orientation are encoded, and therefore allow relating neuronal representation to linear algebra concepts. Classical HD tuning, gravity tuning, and DA/TA can all be described within this common framework. Together, these results provide a new mathematical, geometrical, and topological perspective on 3D HD coding.

## Introduction

The head direction (HD) cell system, also known as the brain's internal compass, is a distributed neural system that tracks head orientation in space (Angelaki & Laurens, 2020; Laurens & Angelaki, 2018; Taube, 2007). Historically, it has been described and studied primarily as a horizontal compass, encoding a single angular degree of freedom: head azimuth in the horizontal plane (**Fig. 1A**, left). Over the past decade, however, a series of studies (Angelaki et al., 2020; Finkelstein et al., 2015; LaChance et al., 2020; Laurens et al., 2016; Page et al., 2018; Williams et al., 2026) has demonstrated that the HD system has fundamental 3D properties, which can be divided into two distinct and complementary aspects of 3D HD coding.

- First, multiple studies (Angelaki et al., 2020; Finkelstein et al., 2015; Laurens et al., 2016) have shown that HD-related brain regions encode head tilt relative to the vertical (**Fig. 1A**, right). Unlike azimuth, which is described by a single angular variable, head tilt has two degrees of freedom and specifies the orientation of the head's vertical axis relative to the allocentric vertical. Subsequent experiments (Angelaki et al., 2020) demonstrated that this representation is anchored to gravity, which is sensed by the otolith organs in the inner ear. Tilt coding has therefore also been referred to as gravity tuning.

- A second line of work has addressed a different question: how is azimuth encoded when the head itself is tilted away from upright? Across laboratories, a common model has emerged, initially proposed as the dual-axis (DA) model (**Fig. 1B**, see legend for definition) and subsequently described, using a different mathematical formulation, as the tilted-azimuth (TA) model (**Fig. 1C**, see legend for definition). Altogether, experimental support for the DA/TA model has been obtained by several laboratories (Angelaki et al., 2020; LaChance et al., 2020; Laurens & Angelaki, 2019; Page et al., 2018; Williams et al., 2026). This framework also generalizes earlier descriptions that were restricted to rotations about particular axes, including the toroidal geometry proposed by Finkelstein et al. (Finkelstein et al., 2015) and earlier observations of HD-cell activity (Taube, 2011; Taube et al., 2013).

Conceptually, gravity tuning is rather straightforward to describe. Angelaki et al. (Angelaki et al., 2020) characterized it as a neuronal population code with Gaussian tuning to the 3D gravity signal. Gravity is sensed by the otolith organs of the inner ear as a 3D vector, and gravitational signals are further extracted from vestibular information by brainstem and cerebellar circuits (Laurens, 2022; Laurens et al., 2013). The orientation and magnitude of the gravity vector relative to the head are therefore readily available to HD-related regions, presumably through direct or indirect projections from the vestibular nuclei to the HD system (Mehlman et al., 2021).

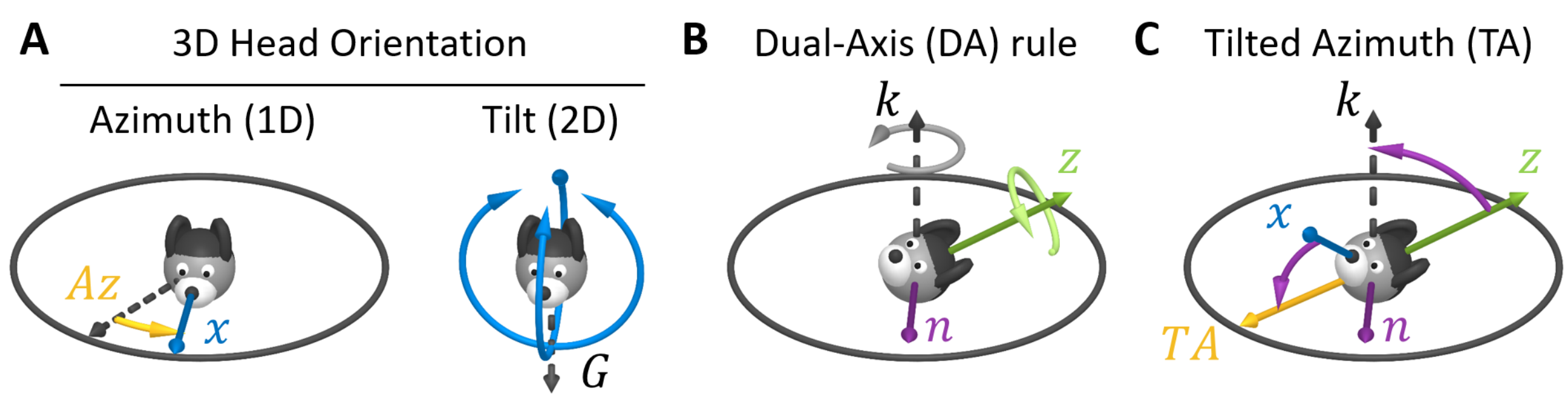

**Figure 1: Current theoretical framework for 3D encoding of head orientation by HDCs. A:** 3D head orientation is decomposed into azimuth, i.e. head orientation in the horizontal plane (typically defined as the direction of the head's forward axis, x), and tilt, i.e. head orientation relative to allocentric vertical, which is anchored to gravitational vertical (G). **B:** Dual-axis rule: 3D rotations are decomposed into a rotation around the head's vertical axis (z) and the hearth-vertical axis (k). The azimuth signal is updated by the sum of these two rotations. Note that, in general, 3D rotations may have a component around a third axis, n, perpendicular to z and k. This component doesn't update azimuth. **C:** Tilted azimuth model: starting from any 3D orientation, TA can be computed by rotating the head is back to upright through the shortest possible rotation. The TA corresponds to the orientation of the head's forward axis following the rotation. Note that the axis of rotation is identical to n, as defined in B. See **Fig. 2** for a complete description of the head's axes and allocentric axes.

In contrast, although the DA/TA model is well defined and strongly supported experimentally, its mathematical formulation is considerably less intuitive. Existing formulations (**Fig. 2B,C**) require complex geometrical transformations and contain a singularity at the inverted head orientation:

when the head is exactly upside down, TA becomes undefined because the definitions in **Fig. 2B,C** admit an infinity of solutions. Furthermore, small head rotations in the vicinity of this orientation can lead to large changes in TA. This raises a general mechanistic question: how might neural circuits actually compute such a representation? A further complication is that the azimuthal tuning of HDCs progressively weakens as the head tilts away from upright and toward inversion. Empirically, this attenuation can be described by a tilt-dependent gain factor, approximating a smooth cosine-like function (Angelaki et al., 2020; Laurens & Angelaki, 2019). Thus, existing descriptions require both a relatively complex transformation to define TA and an additional tilt-dependent modulation of its neuronal representation.

A useful perspective on this problem comes from recent work on the representation of 3D rotations in artificial neural networks. Zhou et al. (Zhou et al., 2019) pointed out that familiar low-dimensional parameterizations of rotation, including Euler angles and quaternions, necessarily introduce discontinuities when represented in Euclidean spaces of four or fewer dimensions. They argued that, for neural networks, it can therefore be advantageous to abandon minimal coordinate descriptions in favor of higher-dimensional, continuous representations of rotation.

Here, I show that this is indeed possible for DA/TA azimuth. Starting directly from a rotation matrix describing 3D head orientation, TA and its tilt-dependent gain can be represented jointly by a simple 2D vector obtained through a linear operation. The direction of this vector specifies TA, whereas its magnitude decreases continuously as the head tilts away from upright, reaching zero when the head is inverted. Thus, DA/TA becomes mathematically simpler when represented in this form, and the singularity at inversion is naturally resolved: azimuth itself remains undefined at the inverted orientation, but the magnitude of its neural representation goes to zero, such that the representation as a whole remains continuous.

Finally, I place this formulation within a more general framework, directly inspired by Zhou et al. (Zhou et al., 2019), where head orientation is described by the nine elements of a rotation matrix, and neuronal responses are modeled using a linear–nonlinear (L–NL) cascade applied to these elements. I focus primarily on the linear stage of this cascade, which defines the information bottleneck of the model—that is, which information about 3D head orientation is encoded by the neuron before the subsequent response nonlinearity is applied. Within this framework, both tilted-azimuth coding and gravity tuning can be expressed using simple linear combinations of rotation-matrix elements, providing a simpler starting point for investigating how these representations could be generated by neural circuits.

## A General Framework for Continuous Coding of 3D Head Orientation

Following (Zhou et al., 2019), we begin from the observation that a 3D rotation can be represented continuously by the nine components of a 3 × 3 rotation matrix. Let (i, j, k) denote a fixed allocentric coordinate frame, and let (x, y, z) denote the three head-fixed axes (**Fig. 2A**). Head orientation can then be described by a rotation matrix $R$ (**Fig. 2B**).

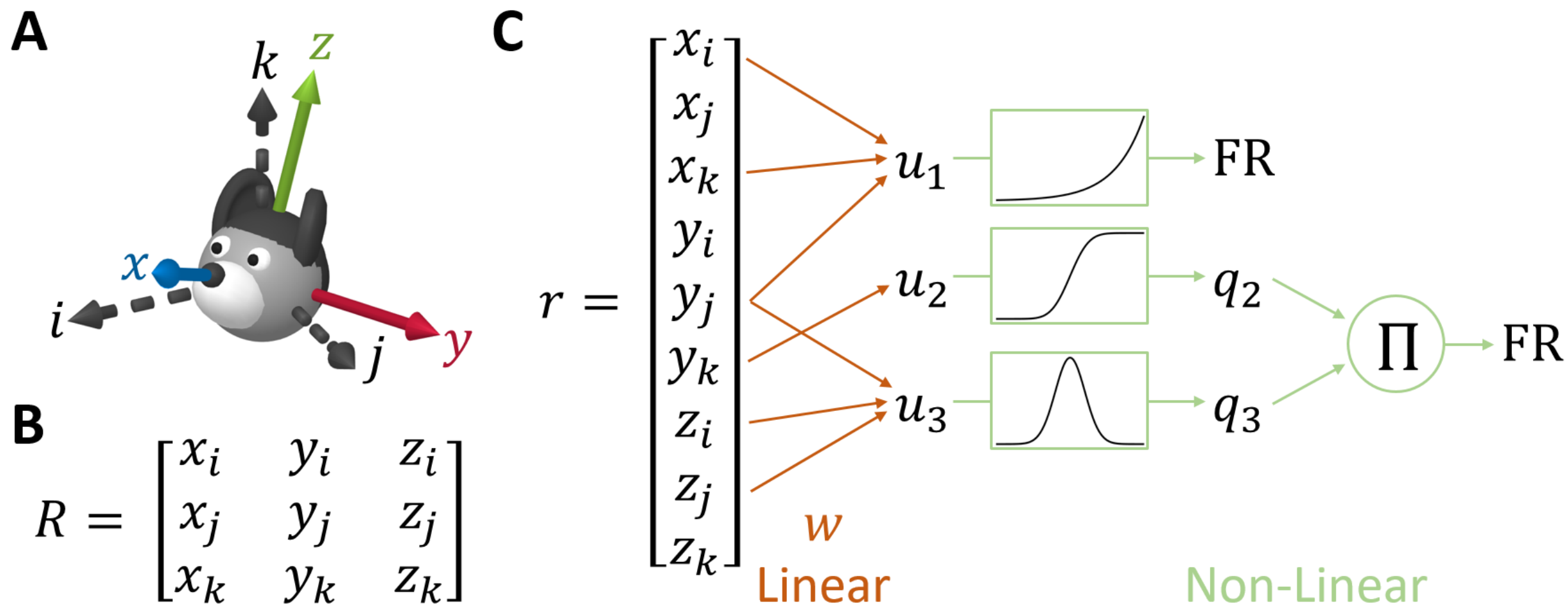

**Figure 2: General framework. A:** representation of the allocentric $(i, j, k)$ and egocentric $(x, y, z)$ reference frames. **B:** associated rotation matrix. Here, for example, $x_i = x \cdot i$ is the projection of the head-fixed $x$-axis onto the allocentric $i$-axis. **C:** structure of a R-L-NL cascade: elements the $R$ are fed though a linear stage, and then a non-linear stage (that may involve a simple non-linear function or combine multiple ones).

Note that the nine entries of $R$ vary continuously as the head rotates. Therefore, one could theorize that a set of neurons encoding these 9 components, e.g. in the form a linear rate code, would provide a simple (and continuous) neuronal code of head orientation.

Yet, as pointed out in (Zhou et al., 2019), this representation is redundant because the rows and columns of a rotation matrix are mutually orthogonal unit vectors. Any one row can be reconstructed from the other two using a vector product, and the same is true for the columns. A six-dimensional representation could therefore be obtained by retaining only two rows or two columns of the matrix, and Zhou et al. provide an additional reduction to 5 dimensions. We do not make this simplification here because the full matrix provides a convenient starting point.

Here, we propose a framework where neural responses are constructed from the 9 components of (R) through three operations: a linear transformation (L) followed by one (or multiple) stage of non-linear operations (NL). Accordingly, we name this framework R-L-NL. For convenience, we first collect the nine components of the rotation matrix into a vector,

$$r = \left(x_i, x_j, x_k, y_i, y_j, y_k, z_i, z_j, z_k\right)^T$$

The first operation is that a neuron may compute a weighted linear combination of these components. For neuron $(n)$, this can be written as

$$u_n = b_n + w_n \cdot r \quad ,$$

Where $w_n$ is a vector containing the weights assigned to the nine components of R, and $b_n$ is a constant offset.

The second operation is that the resulting value may then be passed through a continuous nonlinear function,

$$FR_n = f_n(u_n) \ .$$

Possibly, and without loss of generality, the nonlinear stage may be decomposed into a multi-stage operation, i.e. the product of two nonlinear functions, as illustrated in **Fig. 2C**. It could also represent more complex processes, e.g. attractor dynamics. In all events, we assume that the NL stage is continuous.

Within this framework, the activity of individual neurons (or the population) is a continuous function of head orientation. Note that this framework does not imply that complete head orientation can be uniquely recovered from the population response: that depends on the nature of the linear/non-linear combinations present at the population level.

Note also that it is quite clear that such a framework, that combines arbitrary linear and non-linear function, can be used to fit practically any type of neuron – there is nothing particularly original in that. The point that we will make through this manuscript is that (1) the representation is continuous and (2), provided that the non-linear stages don't lose any information, *the nature of the information represented depends on the first linear stage*. Thus, our focus will be on understanding *what* this first linear stage represents, which in turns will allow to understand the underlying sensory processing using linear algebra concepts.

## Comparison with current models of HDC

First, we show that this framework can include already established models of HD and gravity-related tuning.

## Head direction tuning

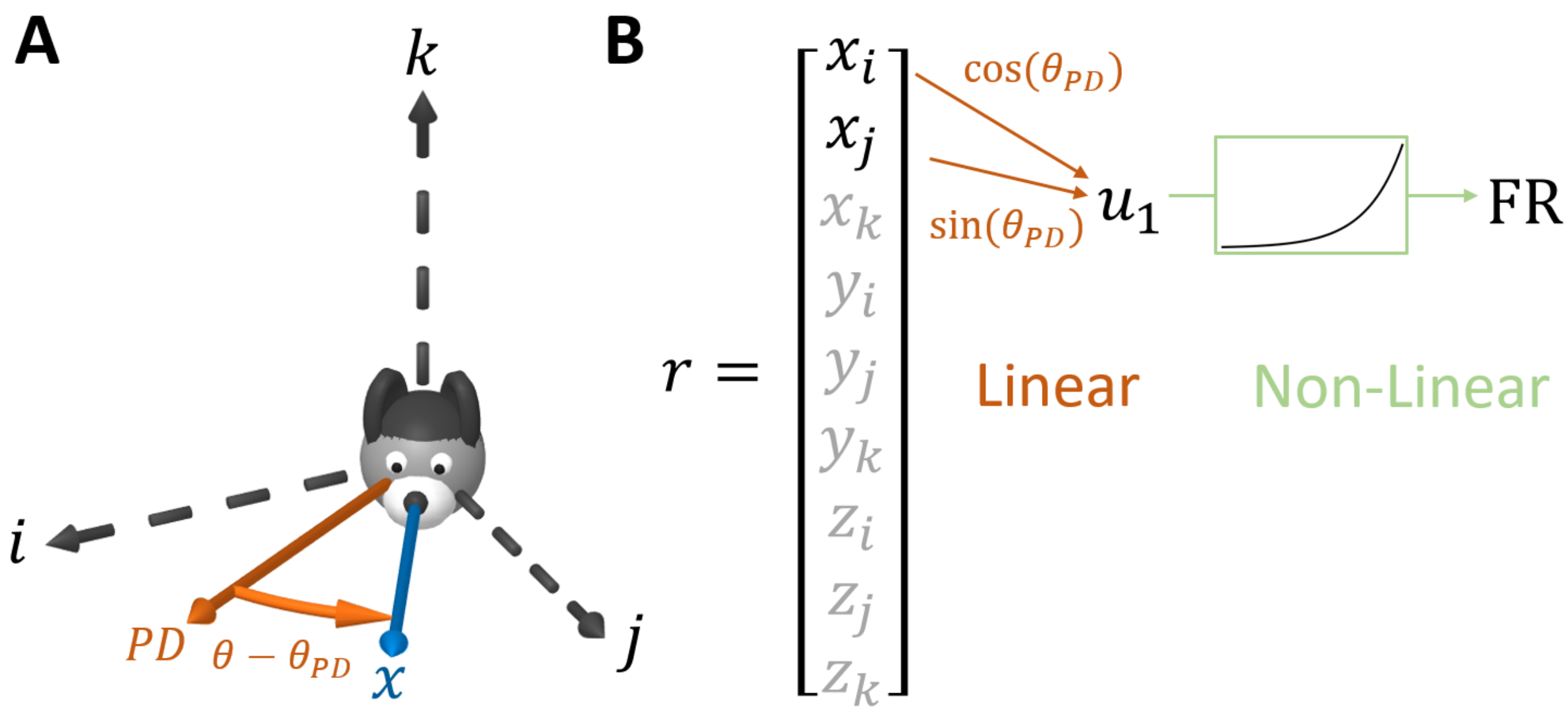

**Figure 3: Encoding of 'classical' azimuth in a R-L-NL cascade. A:** Geometrical illustration of classical azimuth. **B:** Decomposition of a classical HDC tuning curve (i.e. a von Mises) as a R-L-NL cascade.

Consider first a classical HDC recorded while the animal remains upright and rotates only in the horizontal plane. Its firing rate is commonly described by a von Mises function:

$$FR(\theta) = A\, exp[\kappa\, cos(\theta - \theta_{PD})]$$

Here, θ is head direction, $\theta_{PD}$ is the preferred direction, $\kappa$ is a factor controlling tuning width, and $A$ is a scaling factor.

In our framework, $cos(\theta - \theta_{PD})$ can be expressed as a linear combination and the exponential as the non-linear stage, as shown below.

Since we assume that the head is upright, $x_k = 0$, and the horizontal head direction is encoded by the first two components of the head's $x$ axis:

$$x = (x_i, x_j, 0) = (cos\,\theta, sin\,\theta, 0)$$

A cell's preferred direction $(PD)$ can be represented by the unit vector

$$PD = (cos\,\theta_{PD}, sin\,\theta_{PD}, 0)$$

The term $cos(\theta - \theta_{PD})$ can be rewritten as a dot product:

$$cos(\theta - \theta_{PD}) = cos\,\theta_{PD} x_i + sin\,\theta_{PD} x_j = PD \cdot x$$

Hence, the term $cos(\theta - \theta_{PD})$ is a linear combination of $x$. In the general R-L-NL framework, it can be written as $w.r$ with $w = ( cos\ \theta_{PD}, sin\ \theta_{PD}, 0,0,0,0,0,0,0)$.

Next, the exponential term of the von Mises distribution can then be formalized by a the NL stage $f(u) = A\ exp[\kappa.u]$.

Altogether, the von Mises model can therefore be rewritten as a R-L-NL model. Importantly, this model indicates that, as long as the head is horizontal, HDCs encode the vector $(x_i, x_j)$ in a way that can be described as a cosine tuning (in the first linear stage) followed by an exponentiation.

## Gravity Tuning of Head Direction Cells

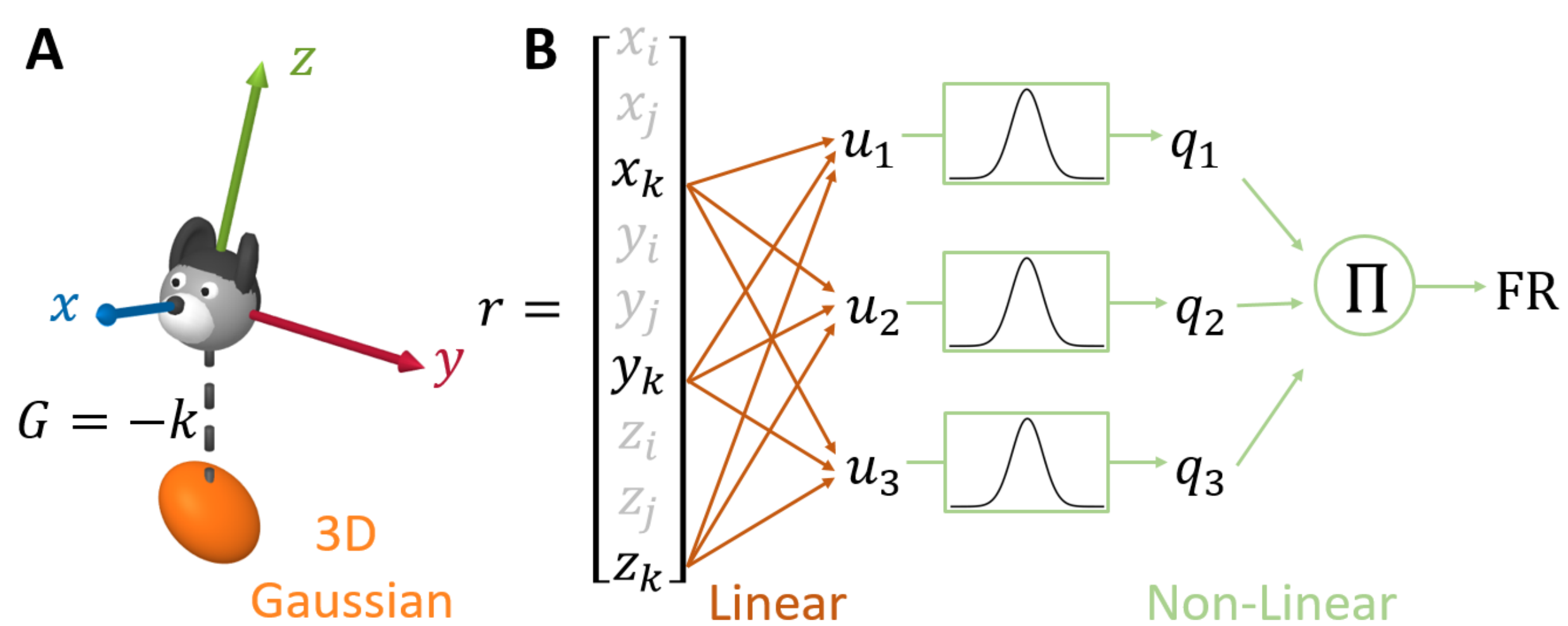

**Figure 4: Encoding of gravity tuning in a R-L-NL cascade. A:** Geometrical illustration of gravity coding as a 3D Gaussian. **B:** Decomposition of a 3D Gaussian tuning curve as a R-L-NL cascade.

The same R–L–NL framework can describe tuning to head tilt relative to gravity. In (Angelaki et al., 2020), gravity-related tuning was modeled as a 3D Gaussian function of $G$, where $G$ is a normalized gravity vector expressed in an egocentric reference frame (**Fig. 4A**).

In allocentric coordinates, the normalized gravity vector is $(0,0,-1)^T$. Expressed in head-fixed coordinates, it is

$$G = R * (0,0,-1)^T = (-x_k, -y_k, -z_k)^T$$

Each component of the egocentric gravity vector is therefore already contained in the rotation matrix (**Fig. 4B**).

Next, a 3D Gaussian can be expressed as a L-NL combination of G, as follows. The matrix equation of a 3D Gaussian is:

$$f(G) = A\,exp\left[-\left(\frac{1}{2}\right)(G-\mu)^T W^T W (G-\mu)\right]$$

where μ is the preferred gravity vector for a given cell, and $W$ determines the shape (principal axes) and width of the 3D Gaussian function. We can define:

$u = (u_1, u_2, u_3) = WG$ and $u_\mu = (u_{\mu,1}, u_{\mu,2}, u_{\mu,3}) = W\mu$

Intuitively, $u_1, u_2$ and $u_3$ are the three principal axes of the Gaussian distribution, scaled by the Gaussian's variance along these axes, and $u_\mu$ is the Gaussian's center along these axes.

And the Gaussian is rewritten as:

$$f(G) = A\,exp[-\frac{1}{2}(u-u_\mu)^T(u-u_\mu)]$$

Or equivalently:

$$f(G) = A\,exp\left[-\tfrac{1}{2}(u_1-u_{\mu,1})^2\right] exp\left[-\tfrac{1}{2}(u_2-u_{\mu,2})^2\right] exp\left[-\tfrac{1}{2}(u_3-u_{\mu,3})^2\right]$$

In this formula, the term. $exp\left[-\tfrac{1}{2}(u_1-u_{\mu,1})^2\right]$ includes $u_1$, which is a linear combination of G (i.e. a linear combination of $R$ since $G$ is a subset of $R$), fed through a Gaussian nonlinearity (with mean $u_{\mu,1}$). The corresponding terms including $u_2$ and $u_3$ have a similar structure, and the three terms are multiplied together (**Fig. 4B**).

In the R-L-NL framework, these operations can be formalized as:

$$u_1 = w_1.r \text{ with } w_1 = (0,0,u_1^x,0,0,u_1^y,0,0,u_1^z)$$

$$q_1 = f(u_1 - u_{\mu,1}) \text{ where } f \text{ is a Gaussian function}$$

$$FR = A\prod_{n=1}^{3} q_n$$

To conclude, although the mathematical description of the 3D Gaussian tuning function is somewhat involved, the underlying representation is simple. Gravity tuning amounts to encoding the direction of gravity in head-centered coordinates. This egocentric gravity vector is directly available from the rotation matrix ($R$), as its last row, $G = (-x_k, -y_k, -z_k)^T$. Individual gravity-tuned cells can then be modeled by applying a 3D Gaussian tuning function to this vector, with each cell characterized by its preferred gravity vector and tuning widths.

# A new expression for Tilted Azimuth

## Background

We have described how the R–L–NL framework accounts for classical models of HDC tuning in 1D, i.e. when the head is upright. During general 3D movements, experimental studies have shown that HDCs follow a more general model, first developed by Page et al. (Page et al., 2018) (under the name Double-Axis rule) and subsequently by Angelaki et al. (Angelaki et al., 2020; Laurens & Angelaki, 2018) (under the name Tilted Azimuth, i.e. $TA$ which I will use here).

In Page et al.'s definition, head rotation ($\delta\Omega$) is be decomposed into a rotation ($\delta\Omega_z$) around the egocentric yaw axis $z$ and a rotation ($\delta\Omega_k$) around the allocentric axis $k$ (**Fig. 1B**), and HDCs are updated by the sum of these components, i.e.

$$\delta TA = \delta\Omega_z + \delta\Omega_k$$

Note that this decomposition is unique except in two orientations, (1) when upright, in which case the axes $z$ and $k$ are aligned, and a rotation $\delta\Omega$ can be expressed as $\delta\Omega_z$ or $\delta\Omega_k$ indifferently; and (2) when upside down, in which case the axes $z$ and $k$ are antiparallel, and in which case $\delta TA$ can't be defined.

I also provide a brief and straightforward extension of this model, by defining a third 'null' rule. In general, any 3D rotation vector will be the sum of 3 components, around that vectors $z$ and $k$ defined above, and the third vector $n$ defined as $n = z \times k$, i.e. perpendicular to $z$ and $k$. This vector can be seen as a 'null' axis, as rotations around it don't update $TA$. Defining it is useful e.g. for conceptualization and comparison with other models.

In Angelaki et al.'s definition (Angelaki et al., 2020; Laurens & Angelaki, 2018), TA is expressed as follows (**Fig. 1C**): starting from a given head orientation, one rotates the head toward upright, by the shortest possible rotation, which is through an axis orthogonal to both $z$ and $k$ (i.e. $n$) until it reaches the upright orientation. The TA at the original orientation is then defined as the conventional head azimuth of this upright orientation.

Both definition are equivalent and can be used to compute TA at any head orientation (except upside-down). Nevertheless, both formulations rely on relatively involved geometric constructions, which complicated efforts to understand how the brain may compute or update TA.

In addition to the TA computation itself, studies in rodents (Angelaki et al., 2020; Calton, 2005; Laurens & Angelaki, 2019) have revealed that HDCs progressively lose their directional tuning as the animal tilts beyond approximately 90°. Some studies have described this loss as an abrupt disappearance of tuning (Calton, 2005), whereas others (Angelaki et al., 2020; Laurens & Angelaki, 2019) have reported a more gradual attenuation, approximately following a cosine function of head tilt. Within existing models, this phenomenon is typically treated as an additional gain modulation applied after the TA has been computed, introducing an extra stage of processing.

## New Expression

Here, I propose a new expression for TA that simultaneously accounts for both the directional signal and its tilt-dependent attenuation, while fitting naturally within the proposed R–L–NL framework. The new expression is based on the two variables

$$u_1 = 1/2\ \left(x_i + y_j\right)$$

$$u_2 = 1/2\ \left(x_j - y_i\right)$$

As shown in Appendix, TA is then given by the direction of the vector

$$u_{TA} = (u_1, u_2)$$

i.e. $TA = atan2(u_2, u_1)$. Furthermore,

$$norm(u_{TA}) = 1/2.(1 + \cos\alpha)$$

Where $\alpha = acos(z_k)$ is the head-tilt angle, defined as the angle between the head's vertical axis and allocentric vertical.

Thus, just as the direction of the vector $x$ in the allocentric horizontal plane corresponds to traditional head azimuth, the direction of the vector $u_{TA}$ encodes TA during general 3D head movements. In addition, the norm of this vector decreases systematically with head tilt: it equals 1 when the head is upright, 1/2 when the head is tilted by 90°, and 0 when the head is fully inverted. The same representation therefore accounts for the progressive attenuation of HDC tuning observed experimentally, without requiring an additional gain-modulation stage.

This expression for TA fits naturally in the R–L–NL framework (see **Fig. 5A**). Consider a HDC with preferred TA $\theta_{PD}$. Similar to the classical azimuth model, we can express the first linear stage as:

$$u = cos\,\theta_{PD} u_1 + sin\,\theta_{PD} u_2$$

Expanding the terms $u_1$ and $u_2$ based on the definitions above, we obtain $u = w \cdot r$ with

$$w = 1/2\,.(cos\,\theta_{PD}, sin\,\theta_{PD}, 0, -sin\,\theta_{PD}, cos\,\theta_{PD}, 0, 0, 0, 0)$$

The neuronal response is then obtained by applying a nonlinear tuning function,

$$FR = f(u)$$

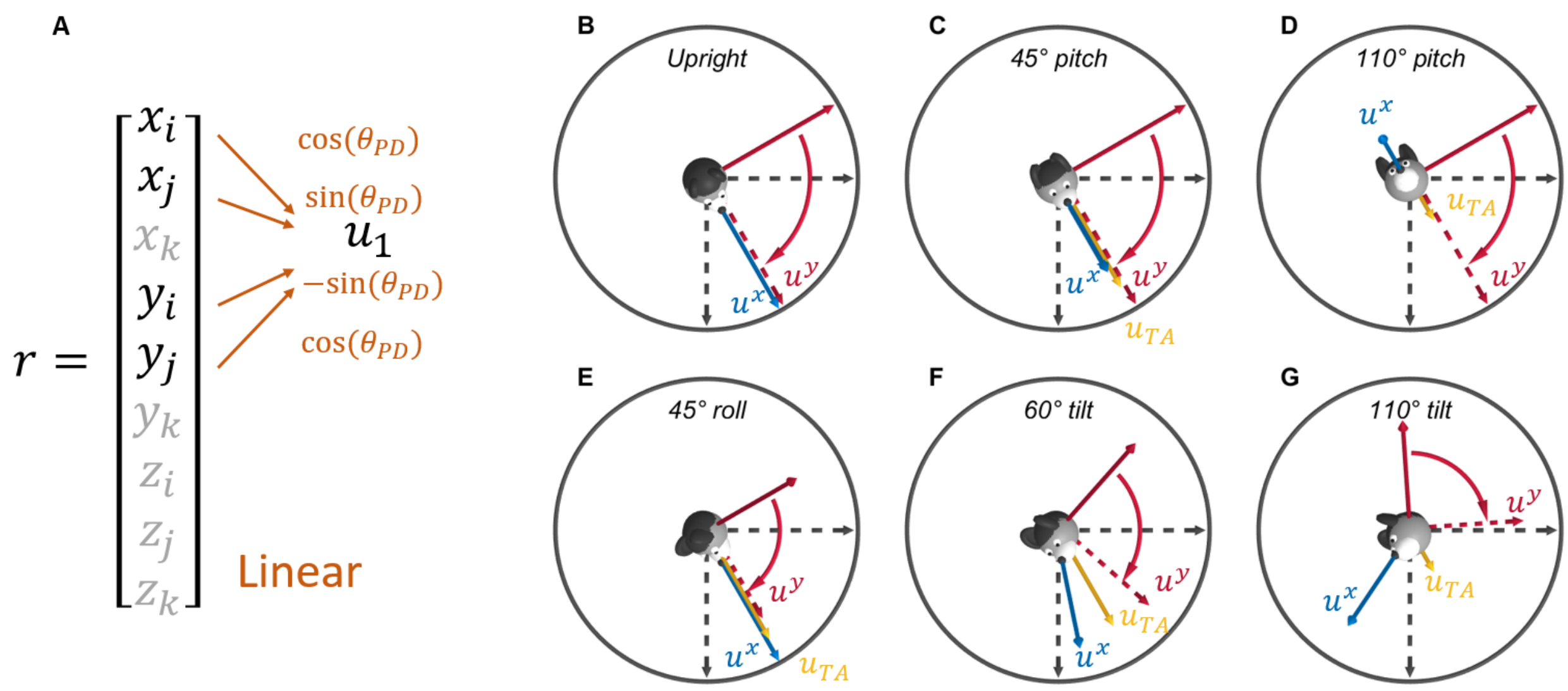

**Figure 5: Encoding Tilted Azimuth in a R-L-NL cascade. A:** Illustration of the linear stage. **B-G:** Geometrical illustration of TA calculation in a variety of head orientations.

## Interpretation

The above formulation also admits a simple geometric interpretation, illustrated in **Fig. 5B-G**. Throughout this section, the allocentric horizontal plane is viewed from above, such that the arrow representing the head-fixed axes $x$ and $y$ (blue and red) directly represent their projection onto allocentric horizontal.

Now, recall that

$$u_1 = (1/2).\left(x_i + y_j\right)$$
$$u_2 = (1/2).\left(x_j - y_i\right)$$

First, suppose that these variables were instead defined as

$$u_1^x = x_i$$
$$u_2^x = x_j$$

The vector $u^x = (u_1^x, u_2^x)$ is simply the projection of the head's forward (nose) axis onto the earth-horizontal plane (i.e. the solid blue arrow in **Fig. 5B**).

Next, consider

$$u_1^y = y_j$$
$$u_2^y = -y_i$$

The vector $u^y = (u_1^y, u_2^y)$ is the projection of the head's left axis ($y$) onto the earth-horizontal plane (**Fig. 5B**, solid red arrow), *rotated by 90° clockwise* (**Fig. 5B,** dashed red arrow).

This construction can be understood most easily in the upright configuration (**Fig. 5B**). Suppose that azimuth were defined relative to the direction of the left ear rather than the nose. This would provide an equally valid measure of heading, but its zero direction would need be shifted by 90°, such that the 'nose referenced' and 'left-ear referenced' azimuth stay consistent. This can also be accomplished by rotating the projected 'left ear' vector by 90° clockwise to obtain $u^y$. As shown in **Fig. 5B**, where the resulting vector (dashed red arrow) matches the 'nose' vector $u^x$ (solid blue arrow).

The TA vector is simply the average of $u^x$ and $u^y$:

$$u_{TA} = (u_1, u_2) = (u^x + u^y)/2$$

When the head is upright (**Fig. 5B**), $u_{TA}$, $u^x$ and $u^y$ are identical. If the head pitches nose-up (**Fig. 5C**), however, the horizontal projection of the nose axis progressively decreases, whereas the rotated ear vector remains unchanged. Because the two vectors remain aligned, their average points in the same direction but has a reduced magnitude, exactly as predicted by the expression for $norm(u_{TA})$.

If the head pitches further until it exceeds 90° (**Fig. 5D**), the direction of $u^x$ reverses. Nevertheless, its magnitude remains smaller than $u^y$, such that the average vector $u$ still points in the same direction as $u^y$. This provides an intuitive explanation for a key property of TA: whereas conventional azimuth reverses beyond 90° pitch, TA remains continuous.

**Fig. 5E** illustrates the complementary situation in which the head rolls about the ($x$)-axis. In this case, it is the 'ear' contribution that progressively decreases, while the 'nose' contribution remains unchanged.

Finally, **Fig. 5F–G** illustrate tilts about axes other than pitch and roll. In these situations, the nose and ear vectors are no longer aligned, and their average therefore becomes intermediate both in direction and in magnitude. This captures the fact that TA angle will significantly differ from traditional azimuth (defined by $u^x$, or if it were defined by $u^y$) during such movements.

In summary, the tilt-modulated TA vector $u_{TA}$ admits a simple geometric interpretation based on three concepts:

- That azimuth may be referenced to multiple head axes, i.e. forward, or lateral, with the appropriate shift of the reference direction (represented here by the rotation of $u^y$).
- That, when the head is tilted, azimuth referenced to a particular head axis can be obtained by a simple projection of that axis onto the earth-horizontal plane.
- That the TA is in fact the vector average of these azimuths, across multiple head axes (here $x$ and $y$).

## The topology of the brain’s representation of 3D motion

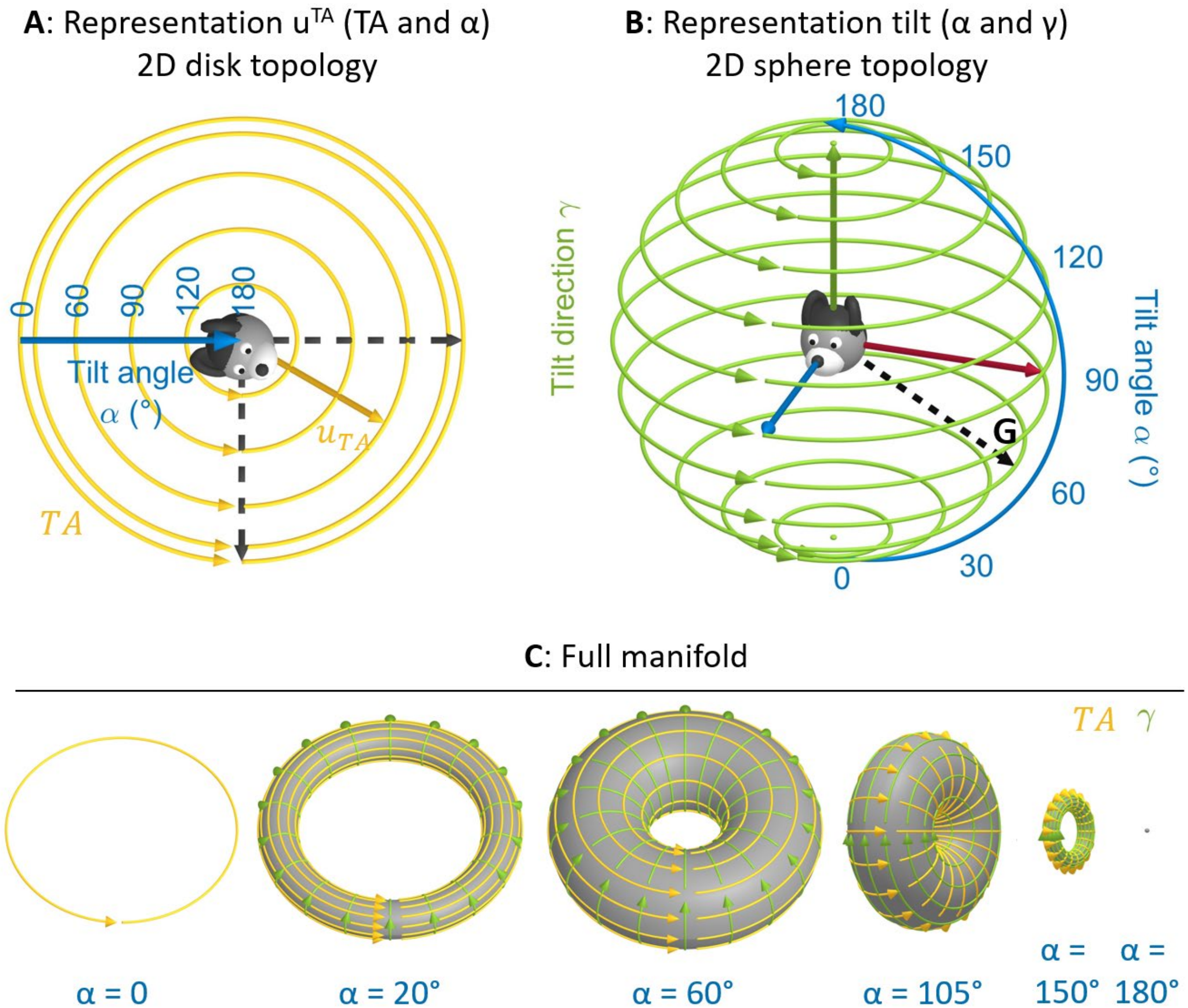

**Figure 6: Topology of the 3D head-orientation representation. A:** Topology of tilt-modulated TA, forming a disk. Each concentric circle corresponds to a fixed head tilt angle relative to vertical ($\alpha$). B: Topology of 2D tilt, forming a sphere. Each parallel on the sphere likewise corresponds to a fixed $\alpha$. **C:** Full manifold. Each panel represents, at a fixed tilt angle $\alpha$, the Cartesian product of a circle in A and the corresponding circle in B, forming a torus. The dimensions of each torus vary according to the radii of the respective circles in A and B. The full manifold is formed by a continuum of these tori.

A fundamental challenge in representing head azimuth arises from the "hairy ball" theorem (Jeffery et al., 2015), which implies that any continuous representation of azimuth over the full range of 3D orientations must contain at least one singularity. In the original formulation of DA of TA, this singularity occurs when the head is upside down. The tilt-modulated TA representation circumvents this problem: it is a continuous representation, devoid of singularity (where the former singularity point is now the origin). However, it sacrifices its ability to encode azimuth in the fully inverted orientation.

The resulting state space of the tilt-modulated TA representation is a two-dimensional disc, within which the vector $u_{TA}$ may assume any value (**Fig. 6A**). Its direction encodes TA (**Fig. 6A,** yellow), whereas its length encodes the absolute head tilt $\alpha = acos(z_k)$ (**Fig. 6A**, blue) in a non-linear manner, i.e. $|u| = (1/2).\left(1 + cos(\alpha)\right)$.

Independently, we previously proposed (Angelaki et al., 2020), and further develop here (**Fig. 4**), that HDCs are also tuned to the components of the gravity vector expressed in egocentric coordinates (**Fig. 6B**, black arrow),

$$G = (-x_k, -y_k, -z_k)^T$$

Because the magnitude of the gravity vector is fixed, ($|G| = 1$) its state space is the surface of the unit sphere (**Fig. 6B**) and can be parametrized by the same angle $\alpha$ (representing the latitude on this sphere) and $\gamma = atan2(-y_k, -x_k)$ (representing the longitude on this sphere, i.e. the direction of head tilt; **Fig. 6B**, green).

How are these two representations related? Each point in the $u_{TA}$ disc specifies a unique combination of TA and tilt angle $\alpha$. On the gravity sphere, the same tilt angle corresponds to a particular parallel of latitude. Thus, to encode 3D head orientation, one could add one circular dimension (encoding $\gamma$) to each point on the disc.

If we now consider a circle on the disc (defined by a fixed tilt angle α and all possible values of TA), adding the dimension γ produces a torus, i.e. the Cartesian product of two circles encoding TA and γ (**Fig. 6C**). As α varies continuously, these toroidal surfaces deform continuously: the radius of the TA circle changes according to $(1 + \cos(\alpha))/2$, while the radius of the γ circle matches the corresponding parallel on the gravity sphere, i.e. $\sin(\alpha)$. At α = 0° (**Fig. 6C**, first panel), the γ circle collapses to a point because the corresponding parallels is a pole of the sphere, and therefore the torus collapses into a single TA disc. At α = 20° (**Fig. 6C**, second panel), the γ circle is now defined (but has a narrow radius) and (TA, γ) now form a torus. At α = 60° (**Fig. 6C**, third panel), the torus becomes thicker as the diameter of the γ circle increases. At α = 105° (**Fig. 6C**, fourth panel), the γ circle is now wider than that formed by TA (forcing us to redraw the torus in another orientation. although the concept remains the same). At α = 150° (**Fig. 6C**, fifth panel), both circles are now very narrow and the torus stats to collapse. Finally, at α = 180° (**Fig. 6C**, last panel), the torus has collapsed onto a single point.

Taken together, these results suggest that the brain's representation of 3D head orientation is organized on a 3D manifold, that can be described as a 1D continuum of 2D toroidal surfaces embedded within the five-dimensional neural state space defined by the variables $(u_1, u_2, x_k, y_k, z_k)$.

## Discussion

Here, I have proposed a general framework for describing the representation of 3D head orientation by HDCs, together with a new formulation of Tilted Azimuth (TA). The central idea, based on (Zhou et al., 2019) is to represent head orientation directly through the elements of a rotation matrix and to consider neuronal responses as resulting from a linear transformation of these elements followed by one or more nonlinear operations. Using this approach, previous models of HDC tuning and gravity tuning can be expressed naturally, while TA admits a particularly simple linear formulation. This formulation jointly represents TA and its tilt-dependent attenuation, provides an intuitive geometrical interpretation, and leads to a continuous representation in which the singularity at the inverted orientation collapses to the origin. Together with gravity tuning, it also suggests a topology for the currently known representation of 3D head orientation.

### R-L-NL framework

The interest of the R–L–NL framework is not in its ability to fit neuronal data per se: there is no doubt that it has enough free parameters to fit a wide range of neuronal response properties. The usefulness in the present context, is to analyze what information is selected as the first linear stage. This is because, provided that the nonlinear stage does not discard information, the linear stage defines what the neurons encode, and allow us to analyze this information through the lens of linear algebra.

Classical HD tuning provides a simple illustration. A von Mises tuning curve can be decomposed into a linear cosine code followed by an exponential nonlinearity. Thus, beneath the familiar bell-shaped tuning curve, the relevant linear representation is simply a cosine code for horizontal azimuth. The same approach also clarifies gravity tuning. Although gravity-tuned responses have previously been modeled using 3D Gaussian functions (Angelaki et al., 2020), the information underlying these tuning curves is considerably simpler: the egocentric gravity vector itself, readily encoded in central vestibular pathways (Laurens, 2022; Laurens et al., 2013).

### A new formulation of Tilted Azimuth

The main new result concerns TA. Previous DA/TA formulations (Angelaki et al., 2020; Laurens & Angelaki, 2018; Page et al., 2018) provide well-defined geometrical rules for updating or computing azimuth during 3D head motion, but these constructions are relatively involved and become singular when the head is exactly inverted. In addition, the experimentally observed attenuation of azimuth tuning with increasing head tilt has been treated as a separate gain modulation (Angelaki et al., 2020; Laurens & Angelaki, 2019). The formulation introduced here combines these two phenomena into a single two-dimensional vector obtained linearly from the rotation matrix. Its direction corresponds to TA, while its magnitude decreases continuously with tilt, from unity when upright to zero when inverted. Thus, TA and its tilt-dependent gain need not be treated as two separate computations: they emerge jointly from the same linear representation.

This formulation also provides a new geometrical interpretation of the DA/TA. Conventional azimuth is typically defined in reference to the head's forward axis, but could also be defined with reference with the head's lateral, or other axes. When the head tilts, each such axis can be projected onto the allocentric horizontal plane to provide an azimuth estimate. The TA vector introduced here can be understood as the average between two such projections – or even more. This interpretation explains both the direction of TA and the decrease in vector magnitude with tilt. Thus, the new expression can be viewed not merely as an algebraic simplification, but as a geometrical description of TA as an average across multiple estimates of azimuth referenced to multiple axes within the head horizontal plane.

### Topology

The state space of the tilt-modulated TA signal is consequently a 2D disc rather than a circle. Combining this disc with the gravity representation provides a possible description of the topology of the full 3D head-orientation code (**Fig. 6**). At each nonzero tilt angle, TA and the direction of head tilt define two circular variables and therefore a toroidal surface. As absolute tilt changes, the dimensions of these tori change continuously. The resulting representation can therefore be viewed as a continuous 3D manifold embedded within the higher-dimensional neural state space defined jointly by the TA and gravity variables. Interestingly, the TA is encoded by two variables and gravity by 3, implying that the neural state space is 5D, as proposed by (Zhou et al., 2019).

An important consequence of this topology of TA concerns the singularity of azimuth at the inverted orientation. The hairy-ball theorem (Jeffery et al., 2015) implies that azimuth cannot be defined continuously over all possible 3D head orientations without a singularity. The present formulation does not make this fundamental problem disappear. Instead, the magnitude of the TA representation decreases continuously to zero as inversion is approached. At the inverted orientation, TA itself remains undefined, but its vector representation has collapsed to the origin and is therefore continuous. In this sense, the singularity is resolved by a loss of azimuth information, and a discontinuity in the neural representation is avoided.

### Tilt-dependent modulation of TA coding

Several studies have reported that the ability of HDCs to encode azimuth decreases when the head tilts. We have described this phenomenon as a smoot, cosine-like gain modulation in (Angelaki et al., 2020; Laurens & Angelaki, 2019) whereas other authors have evoked a more abrupt cutoff beyond 90° tilt (Calton, 2005). The TA vector model described here directly produces a smooth tilt-dependent attenuation, comparable to that observed in our previous studies. However, the model is not incompatible with a more abrupt cutoff, which could result from an appropriate nonlinearity applied to the TA vector. Thus, the precise shape of the attenuation need not be determined by the linear representation itself. In all cases, the central feature of the proposed TA vector, i.e. a progressive loss of azimuth information as the head approaches inversion, is qualitatively consistent with previous reports of tilt-dependent attenuation.

### Coding at the inverted orientation

One particularly interesting feature of this model is therefore the loss of azimuth information at inversion. This property is not introduced uniquely by the present work; it is an explicit feature of the previous DA/TA framework (Angelaki et al., 2020; Laurens & Angelaki, 2018; Page et al., 2018). Taken literally, however, it raises the question of whether an animal can maintain an allocentric heading representation while fully inverted. Existing observations do not yet provide a definitive answer. HD tuning and behavioral performance degrade in rats navigating inverted (Taube, 2011; Valerio et al., 2010), but this may partially reflect the locomotor challenge for rats, which are poorly adapted to inverted locomotion. Conversely, azimuth-related activity has been reported in inverted bats (Finkelstein et al., 2015). One may attribute this distinction to species differences - bats and other highly arboreal species, may be more adapted to navigating inverted. However, I would caution against this conclusion: 3D head orientation should not be equated exclusively with climbing. A terrestrial animal can easily bring the head through very large tilt angles by tilting, lying on their back, etc.

I would therefore argue that further experiments are required to elucidate HD coding in inverted orientation. Should they reveal robust azimuth coding in inverted orientation, this would require either a revision of the DA/TA model itself or the addition of a complementary directional signal, encoded differently.

### Conclusion

What has this work accomplished? First, in my mind, it has clarified the topology of the DA/TA representation, at least based on what we know so far. It has also introduced a new mathematical expression for DA/TA, together with a corresponding geometrical interpretation (**Fig. 5**). I would not necessarily argue that this formulation is superior to the previous ones (**Fig. 1**); rather, I hope that it provides another way to understand this representation, both intuitively and mechanistically.

Since its introduction by Page et al. (Page et al., 2018), the DA/TA rule has struck me as a somewhat counterintuitive; perhaps even bizarre. Yet, a growing body of experimental evidence (including mine, (Angelaki et al., 2020)) supports it. I still find myself confronted with the same fundamental questions: what does this representation mean to the brain? Why does the brain represent head direction in this particular way? Having two distinct formulations of DA/TA (**Fig. 1B,C**) has already given us different ways of looking at the same phenomenon, and therefore different ways of understanding it. Here, I offer a third, together with a geometrical and topological interpretation. It may not yet provide a complete answer to these questions, but I hope that, alongside the work of others, it will help pave the way toward one.

## AI use statement

The manuscript was edited with the assistance of AI software. The novel concepts presented in this manuscript, including the new formulation of Tilted Azimuth (TA), were developed by the author, initially through computer simulations that are not presented here. Instead, I present the resulting formulation together with a mathematical proof developed subsequently. The mathematical demonstration presented in the Appendix was partially generated with AI assistance: the approach was defined by the author, and AI was used to complete the derivation. The resulting demonstration was subsequently verified by the author.

# Appendix

We seek to demonstrate that TA is given by the direction of the vector $u_{TA} = (u_1, u_2)$, i.e. $TA = atan2(u_2, u_1)$, and $norm(u_{TA}) = 1/2 . (1 + \cos(\alpha))$, as defined above, for any head orientation $R$.

## Approach:

Our demonstration is based on the following property of DA or TA azimuth, which is that if, starting from any orientation R, the head is rotated back to upright through the shortest rotation, then TA is equal to the traditional azimuth at that orientation, i.e. $atan2(x_2, x_1)$.

This property is the definition of TA in (Laurens & Angelaki, 2018). We note that this rotation consists in aligning the head axis $z$ with the allocentric axis $k$, and therefore the shortest rotation will be about a vector $n$ perpendicular to both $z$ and $k$. Since it is perpendicular to $k$, $n$ lies in the allocentric horizontal plane.

With this in mind, we can verify that this property is implied by the definition of DA: since $n$ is perpendicular to $k$ and $z$, both the egocentric yaw and the allocentric vertical components of this rotation are zero, and therefore this rotation doesn't change azimuth according to the DA rule. Hence, the azimuth that the initial and at the final upright orientation are equal.

## Demonstration:

### Preliminary notations:

Starting from any matrix R, we define:

$n = z \times k$ as the rotation axis. Since $n$ is earth-horizontal, we write it as $n = (\cos\phi, \sin\phi, 0)$ with $\phi = atan2(-z_i, z_j)$.

$\alpha = -\text{acos}(z_k)$ as the rotation angle from $R^0$ to $R$.

$R^0$ as the matrix at the upright orientation.

$\theta = atan2(\text{x}_\text{j}^0, x_i^0)$ as the TA both at the initial $R$ and upright ($R^0$) orientation.

$x^0 = (\cos\theta, \sin\theta, 0)$ and $y^0 = (\cos(\theta + \pi/2), \sin(\theta + \pi/2), 0) = (-\sin\theta, \ cos\ \theta, \ 0$ the $x$ and $y$ axes at the upright orientation.

### Apply the rotation

We now apply the rotation around $n$ by an angle $\alpha$ to $x^0$ using the Rodriguez formula:

$$x = (x_i, x_j, x_k) = x^0 \cos\alpha + (n \times x^0)\sin\alpha + n\,(n \cdot x^0)(1 - \cos\alpha)$$

We note that $n \times x^0 = (0, 0, \sin(\theta - \phi))$ and $n \cdot x^0 = \cos(\theta - \phi)$

Leading to:

$$x_i = \cos\theta\cos\alpha + \cos\phi\cos(\theta - \phi)\,(1 - \cos\alpha)$$

$$x_j = \sin\theta\cos\alpha + \sin\phi\cos(\theta - \phi)\,(1 - \cos\alpha)$$

Likewise, we apply the rotation around $n$ by an angle $\alpha$ to $y^0$ using the Rodriguez formula:

$$y = (y_i, y_j, y_k) = y^0 \cos\alpha + (n \times y^0)\sin\alpha + n\,(n \cdot y^0)(1 - \cos\alpha)$$

We note that $n \times y^0 = \left(0, 0, \sin\left(\theta + \frac{\pi}{2} - \phi\right)\right) = (0, 0, cos(\theta - \phi))$ and $n \cdot y^0 = \cos\left(\theta + \frac{\pi}{2} - \phi\right) = -\sin(\theta - \phi)$

Leading to:

$$y_i = -\sin\theta\cos\alpha - \cos\phi\sin(\theta - \phi)\,(1 - \cos\alpha)$$

$$y_j = \cos\theta\cos\alpha - \sin\phi\sin(\theta - \phi)\,(1 - \cos\alpha)$$

## Compute $u_1$ and $u_2$

We now compute:

$$u_1 = 1/2\,.\left[x_i + y_j\right]$$

$$u_1 = 1/2\,.[\cos\theta\cos\alpha + \cos\phi\cos(\theta - \phi)\,(1 - \cos\alpha) + \cos\theta\cos\alpha - \sin\phi\sin(\theta - \phi)\,(1 - \cos\alpha)]$$

Using the identity $cos(A + B) = cos\,A\,cos\,B - sin\,A\,sin\,B$ with $A = \varphi, B = \theta - \varphi$:

$$cos\,\varphi\,cos(\theta - \varphi) - sin\,\varphi\,sin(\theta - \varphi) = cos\,\theta$$

$$u_1 = 1/2\,.[2cos\,\theta\,cos\,\alpha + (1 - cos\,\alpha)cos\,\theta] = 1/2\,.[cos\,\theta\,(1 + cos\,\alpha)]$$

Likewise:

$$u_2 = 1/2\,.\left[x_j - y_i\right]$$

$$u_2 = 1/2\,.[\sin\theta\cos\alpha + \sin\phi\cos(\theta - \phi)\,(1 - \cos\alpha) - \cos\theta\cos\alpha + \sin\phi\sin(\theta - \phi)\,(1 - \cos\alpha)]$$

Using the identity $sin(A + B) = sin\,A\,cos\,B + cos\,A\,sin\,B$ with $A = \varphi, B = \theta - \varphi$:

$$sin\,\varphi\;\;cos(\theta - \varphi) + cos\,\varphi\;\;sin(\theta - \varphi) = sin\,\theta$$

$$u_2 = 1/2\,.[2\,sin\,\theta\,cos\,\alpha + (1 - cos\,\alpha)\,sin\,\theta] = 1/2\,.[sin\,\theta\,(1 + cos\,\alpha)]$$

## QFD

Based on the above, we have

$$u_{TA} = (u_1, u_2) = 1/2\,.(1 + \cos\alpha).(\cos\theta\,, sin\theta)$$

Note that $1/2\,.(1 + \cos\alpha)$>0, therefore $u_{TA}$ points in the direction of the TA, i.e. $\theta$; and its norm is $1/2\,.(1 + \cos\alpha)$.